
\input harvmac
\def\gs{g_{\rm st}}
\def\np{Nucl. Phys. }

\def\pr{Phys. Rev. }
\def\prl{Phys. Rev. Lett. }

\def\cs{{\cal S}}

\def \cl{{\cal L}}

\Title{\vbox{\hbox{EFI-93-30}
\hbox{\tt hep-th/9306013}}}
{{\vbox {\centerline{Two Dimensional QCD coupled to Adjoint Matter}
\smallskip
\centerline{and String Theory}
}}}

\centerline{\it David Kutasov}
\smallskip\centerline
{Enrico Fermi Institute}
\centerline {and Department of Physics}
\centerline{University of Chicago}
\centerline{Chicago, IL60637, USA}
\vskip .2in

\noindent
We study $2d$ QCD coupled to fermions in the adjoint representation
of the gauge group $SU(N)$ at large $N$, and its relation to string theory.
It is shown that the model undergoes a deconfinement transition at a finite
temperature (analogous to the Hagedorn transition in string theory), with
certain winding modes in the Euclidean time direction turning tachyonic
at high temperature. The theory is supersymmetric for a certain
ratio of quark mass and gauge coupling. For other values of that ratio,
supersymmetry is softly broken. The spectrum of bound states
contains an infinite
number of approximately linear Regge trajectories, approaching at large
mass $M$, $\alpha^\prime M^2=\sum_i i l_i$ $(l_i\in{\bf Z_+})$. Thus, the
theory exhibits
an exponentially growing density of bosonic and fermionic states
at high energy. We discuss these results in light
of string expectations.

\Date{6/93}
%

\newsec{Introduction}

G. 't Hooft's suggestion
\ref\th{G. 't Hooft, Nucl. Phys. {\bf B72} (1974) 461.}\
to generalize the gauge group
of QCD from $SU(3)$ to $SU(N)$ and study the theory
in a $1/N$ expansion, is widely believed to be the most promising
approach for obtaining quantitative analytical insight into the physics of
confining gauge theory. The (known) properties of mesons in large $N$
gauge theory are in qualitative agreement with those observed in the real
world (see e.g. \ref\co{E. Witten, \np {\bf B160} (1979) 57;
S. Coleman, {\it ``Aspects of Symmetry"}, Cambridge
University Press (1985).});
yet, the large $N$ theory is expected to be significantly simpler than real
QCD, essentially because the bound states become free as $N\to\infty$
(with interactions of order $1/N$). The results of \th\ and other arguments
(see e.g.
\ref\poly{A. Polyakov, {\it `` Gauge Fields and Strings"}, Harwood
Academic Publishers (1987).})
point to a possible relation of the $1/N$ expansion of gauge theory to the
topological expansion of some string theory, with $1/N$ playing the role
of the string coupling $\gs$. Over the years there have been many attempts
to make the correspondence more precise
(see \ref\pol{J. Polchinski, Texas preprint UTTG-16-92, hep-th/9210045.}\ for a
recent
review), but so far no concrete models have arisen. In fact, with an
improvement
of the understanding of unified string theory, it has been realized
that there are some qualitative difficulties with a string description
of QCD (in a $1/N$ expansion):

\noindent 1) All known string theories describe in one way or another
space-time gravity. There seems to be no trace of that in QCD\foot{
Although perhaps it is not completely ruled out. Space-time gravity does
arise in certain $0,1$ dimensional large $N$ matrix models that received
a lot of attention recently \ref\gm{P. Ginsparg and G. Moore,
1992 TASI lectures, hep-th/9304011.}.}.

\noindent 2) There is a very general relation (to be elaborated on below)
in string theory between the number of physical states and infrared
stability \ref\ks{D. Kutasov and N. Seiberg, \np {\bf B358} (1991) 600.}.
That relation seems, in general, to be violated by gauge theories.

At the present state of the subject
it is difficult to address these and other issues regarding large $N$
gauge theories
in four dimensions. The purpose of this note is to discuss
the situation in two dimensional gauge theory, with the hope of generating
new ideas applicable in four dimensions.
The choice of two dimensions is of course
due to tractability, but we would like to retain as many of the essential
features of $4d$ gauge theory as possible. In particular, it seems that the
number of degrees of freedom (and the related issue of the existence
of deconfinement/Hagedorn transitions) plays an important role in string-like
models. In gauge theory,
much of the non-trivial structure
of \th\ seems to be related to the existence
of propagating fields in the adjoint
representation of $SU(N)$, the transverse gluons.
Thus, pure $2d$ QCD, which has been recently discussed from a string
theory point of view \ref\dg{D. Gross, Princeton preprint
PUPT-1356 (1992), hep-th/9212149;  D. Gross and W. Taylor, preprint
PUPT-1376, LBL-33458, hep-th/9301068.}\ is inappropriate for our purposes,
having no field theoretic degrees of freedom (and no deconfinement transition).
The 't Hooft model \ref\tho{G. 't Hooft, \np {\bf B75} (1974) 461.}\
of complex fermions in the fundamental representation of $SU(N)$ exhibits
a single asymptotically linear meson ``Regge trajectory", and it should be
interesting to reformulate it as a (perhaps open and closed) string theory.
It too does not exhibit a deconfinement transition. We will concentrate
here on the theory of (Majorana Weyl) fermions in the adjoint representation of
the gauge group, described by the Lagrangian,
\eqn\a{\cl={1\over g^2} F_{\mu\nu}^2+\bar \psi\gamma^\mu D_\mu
\psi+m\bar\psi\psi}
where, as usual, $F_{\mu\nu}=\partial_{[\mu}A_{\nu]}+[A_\mu, A_\nu]$,
$D_\mu\psi=\partial_\mu\psi+i[A_\mu, \psi]$, $\psi_{ab}$ is a traceless
hermitian anticommuting matrix, $m$ is the (bare) fermion mass and $g$
the gauge coupling\foot{This model has been recently studied in
\ref\kl{S. Dalley and I. Klebanov, \pr {\bf D47} (1993) 2517.}.}.
The hope is that the adjoint fermions mimick the effect of transverse
gluons and make the theory non-trivial.
The main reason for our interest in \a\ is that (as we'll see
below) this theory
exhibits an infinite number of asymptotically linear ``Regge trajectories",
and undergoes a deconfinement transition at a finite temperature
much like (what is expected in) four dimensional gauge theories.
Nevertheless, it is more amenable to quantitative analysis, which can perhaps
shed light on the relation to string theory and teach us
other valuable lessons about higher dimensional gauge theory.

In section 2 we discuss a method, due to J. Polchinski, to study the
stability of the confining phase at high temperature in a language
familiar in string theory -- in terms of masses of certain winding
modes around the (Euclidean) time direction. Applying these ideas
to $2d$ QCD provides a qualitative guide to the different theories.
We show that models with adjoint matter lose confinement at a finite
temperature, unlike those with fundamental matter or no matter at all.
This implies a rich spectrum of bound states.

In section 3 we turn to a more detailed analysis of QCD coupled
to quarks in the adjoint representation of $SU(N)$. We describe the
light cone quantization of the model, and show that it exhibits
an in general explicitly broken supersymmetry, which is restored
for a certain ratio of quark mass to gauge coupling.

In section 4 we discuss
the spectrum of heavy bound states of adjoint quarks.
We find a spectrum of masses which looks quite ``stringy'',
consisting of an infinite number of asymptotically linear
Regge trajectories. The density of single particle states increases
exponentially with energy.

Section 5 contains some remarks on the topological
space-time theory obtained in the far infrared, and a proof that supersymmetry
is not spontaneously broken in this model. We conclude in section
6 with some
comments about the results and their relation to string theory.

\newsec{The Hagedorn transition in $2d$ QCD}

At finite temperature one studies the theory \a\ on a Euclidean
cylinder $R\times S^1$; the $S^1$ describes Euclidean time
which is periodic:
$\tau\sim \tau+\beta$ ($\beta$ is the inverse temperature), with boundary
conditions for the gauge field and fermions (see e.g.
\ref\gpy{D. Gross, R. Pisarski and L. Yaffe, Rev. Mod. Phys. {\bf 53}
(1981) 43.}):
\eqn\b{A^\mu(\beta, x)=A^\mu(0,x);\;\;\psi(\beta,x)=-\psi(0,x)}
The phase structure is analyzed in terms of the behavior of Wilson loops
winding around the (compact, Euclidean) time direction:
\eqn\c{\rho_k(x)={1\over N} Tr P \exp\left[i\int_0^{k\beta}d\tau
A_0(\tau, x)\right]}
$P$ denotes path ordering; $k$ is the winding number of $\rho_k$. In the
confining phase, correlators of $\rho_k(x)$ fall off exponentially
at large $x$:
\eqn\e{\langle\rho_k(x)\rho_{-k}(0)\rangle\sim \exp\left[-M_k(\beta) x\right]}
At low temperatures ($\beta\to\infty$)
one expects an area law, $M_k^2(\beta)\simeq k^2\beta^2$.
As the temperature is raised, $M_k^2(\beta)$ decrease, until at some
$\beta_c$ the first of the winding modes becomes tachyonic; equivalently,
the effective string tension vanishes (the transition may actually
be first order
\ref\aw{J. Atick and E. Witten, \np {\bf B310} (1988) 291.}\ and so occur
before $\beta_c$ is reached).
Above the critical temperature confinement is lost.

This description of the deconfinement transition is reminiscent of the
standard picture of the Hagedorn transition in string theory
(see e.g. \aw). The similarity has motivated a beautiful suggestion
by J. Polchinski
\ref\jp{J. Polchinski, \prl {\bf 68} (1992) 1267.}\
of a way to probe the stability of the confining
phase in gauge theory by calculating $M_k(\beta)$ \e\ at high temperature,
where perturbative techniques are reliable. We will next outline
the application of these ideas to $2d$ QCD (a more detailed
account will appear elsewhere
\ref\dk{D. Kutasov, to appear.}).

At finite temperature one may choose an analog of the $A_0=0$ gauge:
\eqn\d{ A^0_{ab}(\tau,x)={\theta_a(x)\over \beta}\delta_{ab}}
{}From \c, \d\ it is clear that $\theta_a$ are periodic with period $2\pi$.
The gauge field $A^1_{ab}$ appears quadratically in the action \a\
and can be integrated out,
giving
rise\foot{In the process of integrating out $A_1$
one also finds a $\log\det(\partial_\tau^2+\theta^2)$, which can be shown to
be unimportnat for what follows.}
to an effective action for $\theta,
\psi$:
\eqn\f{\cl_{\rm eff}={1\over g^2\beta^2}(\theta_a^\prime)^2
+\bar\psi\gamma^\mu D^{(0)}_\mu\psi+m\bar\psi\psi+g^2J_1\Omega^{-2}J_1}
where the following notation has been used:
$\partial_x$ is denoted by a prime (here and below),
$D^{(0)}$ is the covariant derivative \a\ with respect to the
gauge field $A_1^{ab}=0$, $A_0^{ab}$ given by \d. $J_1$ is the space component
of the fermion $SU(N)$ current, $J_\mu^{ab}=\bar\psi^{ac}\gamma_\mu\psi^{cb}$,
and $\Omega_{ab} $ is the operator:
\eqn\g{\Omega_{ab}=i\delta_{ab}\partial_\tau+{\theta_{ab}
\over\beta};\;\;\;
\theta_{ab}\equiv \theta_a-\theta_b}
To determine the behavior of $\rho_k$ \c, \e\ we have to integrate
out $\psi$ and study the resulting action for $\theta_a$. This is particularly
easy to do at high temperature, since then the effective super-renormalizable
couplings $g,m$ are small and it suffices to sum one loop diagrams.
Thus we drop the last term in \f\ (we'll keep the mass term for the time being
since its presence will not complicate the calculation and allow us to make
a few points later), and integrate out $\psi$.
The action for $\theta$ we obtain is:
\eqn\h{\cs_\theta={1\over g^2\beta}\int dx (\theta_a^\prime)^2+V(\theta_{ab})}
where $V(\theta_{ab})=-\log\det\left[(D^{(0)})^2+m^2\right]$ and
$\theta_{ab}$ is as in \g.
The potential $V$ is easily evaluated
(for slowly varying $\theta_{ab}$, which is the
case appropriate for high temperature) following \gpy\ (see also
\ref\nw{N. Weiss, \pr {\bf D24} (1981) 475;
{\bf D25} (1982) 2667.}). Dropping a $\beta$ independent constant corresponding
to the zero temperature determinant, one finds:
\eqn\j{V(\theta)={\beta L\over2\pi}\sum_{k=1}^\infty (-)^k
\int_0^\infty {ds\over s^2} \exp\left(-{k^2\beta^2\over4s}-sm^2\right)
\cos k\theta}
$L$ is the length of space.
In the UV limit $\beta\to 0$ the mass $m$ is irrelevant.
Indeed, rescaling $s\to s\beta^2k^2/4$ in \j\ we see that the
effective mass is $m\beta k/2$, as mentioned above.
Thus, as $\beta\to0$ the effective action for $\theta$ approaches:
\eqn\k{\cs_\theta=\int dx \left[{1\over g^2\beta} (\theta_a^\prime)^2
+{2\over\pi}\sum_{a,b=1}^N\sum_{k=1}^\infty (-)^k{1\over k^2\beta}
\cos k\theta_{ab}\right]}
Of course, the last term in \f\ which we neglected (as well as the
mass and other terms) will induce corrections which become significant at
finite $\beta$.

At large $N$ it is convenient to describe the dynamics of $\theta_a$
(or $\rho_k$ \c) by the density:
\eqn\l{\rho(\theta, x)={1\over N}\sum_a\delta(\theta-\theta_a(x))}
In terms of $\rho$, the action \k\ takes the form:
\eqn\m{\cs_\theta={N^2\over\beta g^2 N}\int dx\int d\theta
{1\over\rho(\theta,x)}(\partial_\theta^{-1}\rho^\prime)^2+
N^2\int d x\int\int d\theta_1 d\theta_2\rho(\theta_1)\rho(\theta_2)V(\theta
_{12})/L}
To study the stability of the confining phase at $\beta\to0$ one now assumes
\jp\ that the eigenvalue distribution is approximately uniform, i.e.:
\eqn\n{\rho_{\rm cl}={1\over 2\pi}}
Expanding about $\rho_{\rm cl}$:
\eqn\o{\rho(\theta, x)={1\over 2\pi}\left[1+\sum_{k\not=0}\rho_k(x)
e^{-ik\theta}\right]}
should show whether small fluctuations destabilize \n.
As is suggested by the notation, $\rho_k$ in \o\ are the same as in \c,
so we are actually calculating $M_k(\beta\to 0)$ \e. Plugging
\o\ in \m\ and keeping only quadratic terms we find:
\eqn\p{ \cs_{\rm lin}=N^2\sum_{k\not=0}\int dx\left(
{1\over \beta g^2N k^2}(\rho_k^\prime)^2+{2\over\pi}(-)^k{1\over k^2\beta}
\rho_k^2\right)}
Comparing to \e\ we read off the masses of the winding states:
\eqn\q{M_k^2(\beta\to0)={2g^2N\over\pi}(-1)^k}
We see that, since the odd $k$ winding states are tachyonic,
the confining phase is unstable at high temperature, and deconfinement
must occur long before \q\ is reached. As discussed in \jp, \q\ should
be thought of as an analytic continuation of the confining phase into the
plasma phase, similar to those routinely performed in string theory.
Other comments:

\noindent 1) Eq. \q\ should be compared to Polchinski's result for pure
four dimensional gauge theory:
\eqn\r{M_k^2(\beta\to 0)=-{2 g^2N\over\pi^2\beta^2 k^2}}
Both imply deconfinement, but \q\ seems (at least naively) more like what one
would expect in unified string theory. There, one has generically:
\eqn\s{\alpha^\prime M_k^2=-C+\beta^2 k^2}
valid for all $\beta, k$. $C$ is a universal constant which measures the
exponential growth of the density of states.
While \s\ looks more like \q\ than \r\ (approaching a universal
constant as $\beta\to0$), at finite $\beta$ \q\ and \s\ probably differ
significantly,  at least for generic $m$ for which \j\ ($K_1$ below is a
modified Bessel function):
$$M_k^2(\beta)={2g^2Nmk\beta\over\pi}(-1)^k K_1(mk\beta)
+O(g^4)$$

\noindent 2) The $(-)^k$ in \q\ looks at first sight embarrassing. It is not
clear why, physically, states with even winding in \c\ should behave
differently than those with odd winding. We don't have a good physical
interpretation
of this sign, but
would like to point out that a similar phenomenon occurs in
string theory.

Indeed, the origin of the $(-)^k$ in \q\ is
the antiperiodic boundary conditions imposed on the fermions \b.
In superstring theory, precisely the same antiperiodicity leads to a similar
strong even-odd difference. One way of seeing that is to notice
that antiperiodicity of space-time fermions and modular
invariance lead to the fact that the GSO projection imposed for even
winding $k$, is the same as in the zero temperature theory (in particular
projecting out all tachyons), while for odd $k$ it is {\it opposite}
to that of the zero temperature theory, allowing tachyons above the
Hagedorn transition \aw.
Therefore the constant $C$ in \s\ actually depends on the parity of
$k$ for superstring models. It is positive for odd $k$, and nonpositive
for even $k$.
Another, closely related way of arriving at the
same conclusion is to write the vertex operator of a space-time fermion
momentum state satisfying \b, and show that odd winding Neveu-Schwarz
(bosonic) states are mutually local with respect to it, whereas
even winding states are not, and have to be modified \dk.
The advantage of the second derivation is that it does not use modular
invariance, which may or may not hold for the QCD string (see below).

Thus, we see that the $(-)^k$ in \q\ is consistent with a description of the
model in terms of a NSR string (with a chiral GSO projection). In fact, the
universality of the difference between even and odd winding states
in string theory would have been a strong argument {\it against} a
string interpretation had we not found the above effect in \q.

\noindent 3) We should mention the results of a similar analysis
applied to other models of $2d$ QCD. For pure QCD one finds as $\beta\to 0$:
$M_k^2=0$. Hence, there is no instability; the confining phase persists
at all temperatures.
Models with (any number $N_f$ of ) fundamental representation fermions
give instead of \p:
\eqn\t{ \cs_{\rm fund}={N^2\over\beta g^2N}\sum_{k\not=0}{
1\over k^2}\int dx\left(
(\rho_k^\prime)^2+(-)^k{2N_f\over\pi N}g^2N
\rho_k\right)}
Here too, the confining phase is stable for all temperatures.
For bosonic adjoint matter one finds
$M_k^2(\beta\to0)=-{2\over\pi}g^2N$; it too exhibits a deconfinement
transition. This model
contains no space-time fermions and the lack of the $(-)^k$
compared to \q\ is consistent with a bosonic
string theory (or e.g. a NSR string with a non-chiral GSO projection).

\noindent 4) The deconfinement transition found for adjoint matter is
an indication of an exponentially rising density of single particle
states at high energy.
We will verify this expectation explicitly below. For the other
models mentioned in 3), which confine at all temperatures, the single
particle bound state spectrum is known, and it indeed does not exhibit
an exponential growth in the density of states with energy.

\newsec{Supersymmetry in $2d$ QCD}

The simple calculations in the previous section provide us with some
useful qualitative information about the different models, but to study
the bound state dynamics, a more detailed analysis is needed.
We will review the formalism in this section, whose main purpose is
to establish the notation
and to discuss a certain accidental
supersymmetry which appears in the adjoint model.

It is convenient to use light cone quantization \th, \ref\lc{
H. Pauli and S. Brodsky, \pr {\bf D32} (1985) 1993, 2001;
K. Hornbostel, S. Brodsky and H. Pauli, \pr {\bf D41} (1990) 3814.},
which has been recently applied to this model in
\kl.
{}From now on we denote by
$\psi_{ab}$ the right moving fermions and by $\bar\psi_{ab}$ the left moving
ones.
The $SU(N)$ currents, $J^+_{ab}=\psi_{ac}\psi_{cb}$, $J^-_{ab}=
\bar\psi_{ac}\bar\psi_{cb}$ form right and left moving
level $N$ affine Lie algebras, respectively. In the
gauge $A_-^{ab}=0$ the Lagrangian \a\ takes the form:
\eqn\u{\cl={1\over
g^2}(\partial_-A_+)^2+i\psi\partial_+\psi+i\bar\psi\partial_-
\bar\psi-2im\bar\psi\psi+A_+J^+}
The equations of motion for $A_+$, $\bar\psi$ do not involve
derivatives with respect to $x^+$, the
light cone ``time"; it is easy to integrate them out to obtain
an action solely in terms of the right moving fermions $\psi$:
\eqn\v{\cl_\psi=i\psi\partial_+\psi+g^2J^+{1\over\partial_-^2}J^++
im^2\psi{1\over\partial_-}\psi}
Quantization on constant $x^+$ surfaces gives rise to the
momentum operator:
\eqn\w{\eqalign{P^+=&i\int dx^-\psi\partial_-\psi\cr
                P^-=&\int dx^-\left(im^2\psi{1\over\partial_-}\psi-g^2
J^+{1\over\partial_-^2}J^+\right)\cr}}
Expanding $\psi(x^+=0)$ in modes:
\eqn\x{\psi_{ab}(x^-)={1\over2\sqrt\pi}\int_{-\infty}^\infty
dk\psi_{ab}(k) e^{-ikx^-}}
and imposing the canonical anticommutation relation,
$\left\{\psi_{ab}(x^-), \psi_{cd}(0)\right\}
=\half\delta(x^-)\delta_{a,d}\delta_{c,b}$
we find the mode anticommutation relations:
\eqn\z{\left\{\psi_{ab}(k), \psi_{cd}(k^\prime)\right\}
=\delta(k+k^\prime)\delta_{a,d}\delta_{c,b}}
$\psi_{ab}(k)$ with $k\le0$ are creation operators, whereas the ones with
$k\ge0$ are annihilation operators. The light cone vacuum is chosen
such that:
\eqn\aa{\psi_{ab}(k)|0\rangle=0\;\;\;\;\;\forall k\ge0}
The momentum operators \w\ are normal ordered in the standard fashion,
and take the form:
\eqn\ab{\eqalign{P^+=&\int_0^\infty dk k\psi_{ab}(-k)\psi_{ba}(k)\cr
                 P^-=&m^2\int_0^\infty{dk\over k}\psi_{ab}(-k)\psi_{ba}
(k)+g^2\int_0^\infty{dk\over k^2} J^+_{ab}(-k)J^+_{ba}(k)\cr}}
where $J_{ab}^+(k)$ is given by (for $k\not=0$):
\eqn\ac{J^+_{ab}(k)=
\int_{-\infty}^\infty dp \psi_{ac}(p)\psi_{cb}(k-p)}
$J_{ab}^+(0)=\int_0^\infty dp\psi_{ac}(-p)\psi_{cb}(p)$
must annihilate
all physical states (due to confinement).

Physical states
are obtained by acting with raising operators $\psi_{ab}(-k)$ on the vacuum
\aa. $P^+$ is diagonal in this basis; therefore, eigenmodes
of $P^-$ are also eigenmodes of the mass operator, $M^2\equiv 2P^+P^-$.
In the next section we will address the problem of diagonalizing
$P^-$, but before getting to that it is perhaps useful to point out
an interesting property of this system, namely its supersymmetry.
The original action \a\ doesn't appear to be supersymmetric. In terms
of on shell degrees of freedom, there are no bosonic ones (no gluons
in two dimensions), while the fermionic sector contains $N^2-1$
left and
right moving degrees of freedom. Consider first the special case $g=m=0$.
Then we have a conformal field theory of fermions in the adjoint
representation of $SU(N)$; this system is well known to be superconformal.
The supersymmetry generator is\foot{This is a special case of the appearance
of superconformal symmetry for free fermions $\psi_i$ in the adjoint
representation of any group. The supercurrent is in general
$G(z)=f^{ijk}\psi_i\psi_j\psi_k$, where $f^{ijk}$ are the structure
constants of the appropriate Lie algebra.}:
\eqn\ad{G_0={1\over3}\int dx^-\psi_{ab}\psi_{bc}\psi_{ca}}
The supersymmetry transformation:
\eqn\ae{\eqalign{\delta\psi_{ab}(p)=&\epsilon\{G_0,\psi_{ab}(p)\}=\epsilon
J_{ab}(p)\cr
\delta J_{ab}(p)=&\epsilon[G_0, J_{ab}(p)]=-\epsilon N p\psi_{ab}(p)\cr}}
acts non-linearly on $\psi$.
In general, adding a mass term and/or gauge interaction destroys the
symmetry \ae. But a look back at \v\ reveals that for a particular
value of $m/g$ the interacting Lagrangian is symmetric under \ae. Indeed,
using the (anti) commutation relations \ae\ it is easy to verify
that:
\eqn\af{[G_0, P^-]=0,\;\;\;{\rm for}\;\; m^2=g^2N }
As is standard in conformal field theory, we also have:
\eqn\ag{G_0^2=NP^+}
Hence, $G_0$ commutes with $P^\pm$ and therefore with the mass operator
$M^2=2P^+P^-$. For any state $|{\rm phys}\rangle$ with mass $M$,
$G_0|{\rm phys}\rangle$ is a state of the same mass and the same value
of $P^\pm$, but opposite statistics. Note that $G_0|{\rm phys}\rangle$
can not vanish since by \ag\ $G_0^2|{\rm phys}\rangle=Np^+|{\rm phys}\rangle$
which can never vanish if $M\not=0$. The physical states are exactly paired
by the interaction for the above value of $g/m$.

There is one subtlety in the above discussion that needs to be
addressed\foot{I thank E. Martinec
for discussions on this issue.}:
supersymmetry could be spontaneously broken in this model, since the Witten
index vanishes. Indeed, consider the $g=m=0$ theory on a cylinder
with periodic boundary conditions for $\psi_{ab}$. The central charge
of the theory is $C=\half(N^2-1)$, whereas the dimension of the (Ramond)
ground state is $\Delta_0={1\over16}(N^2-1)$; since $\Delta_0>C/24$,
${\rm Tr}(-)^F=0$. Thus, even when the interaction Lagrangian in \v\
is exactly supersymmetric, in general supersymmetry can be spontaneously
broken. There are many examples where this is known to occur
\ref\wit{E. Witten, \np {\bf B202} (1982) 253.}.
When we discuss the bound states of the Hamiltonian \w\ below we will
return to this issue and show that supersymmetry is in fact not spontaneously
broken in our model (of course, for generic values
of $g/m$ it is explicitly broken). Qualitatively, this is because the theory
is massive, so a Goldstone fermion can not appear.

It is interesting to compare the softly broken supersymmetry of the
adjoint QCD model to string expectations. In \ks\ it is shown that any
infrared stable string theory has the following property:
the quantity
\eqn\ah{f(s)=s\sum_n(-)^{F_n}\int d^dp\exp\left[-s(p^2+M_n^2)\right]}
where the sum runs over the spectrum of single particle states with masses
$M_n$ and fermion numbers $F_n$ (which are free as $\gs\to0$),
must satisfy the relation:
\eqn\finite{\lim_{s\to0}f(s)={\rm finite}}
Eq. \finite\ is roughly equivalent to the statement that IR stable
string theories must have the number of states of a two dimensional
theory with a finite number of fields \ks. In string theory this is achieved by
quite remarkable cancellations between the contributions to \ah\ of bosons
and fermions. There are very few assumptions that go into the proof
of \finite, the main non-trivial one being modular invariance. As mentioned
in the introduction, this universality makes it interesting to examine
the issue in gauge theory in general, and in our example in particular.

At first glance it seems that
in discussing a two dimensional field theory such as $2d$ QCD,
there is not much content
in \finite. That is not the case, because of the very rich spectrum of
states that will be discussed in the next section. We will find
$\simeq \exp(b\sqrt{n})$
(with a certain constant $b$) states with $M^2\simeq n$, making
$f(s\to 0)$ a difference of two highly divergent quantities:
$f_B(s)\sim \sum_ne^{b\sqrt{n}-sn}\sim \exp ({b^2\over4s})$, and similarly
for $f_F(s)$. Of course, for $m^2=g^2N$ the spectrum is exactly supersymmetric,
and $f(s)=f_B(s)-f_F(s)=0$ for all $s$. What is more interesting is that
for {\it generic} adjoint quark mass $m$, when supersymmetry is explicitly
broken, \finite\ still may be satisfied. This is because the soft SUSY
breaking in \v\ splits the low lying states, but asymptotically, at high
mass levels, SUSY is effectively restored. It is interesting that a similar
asymptotic SUSY is precisely what is required in general for infrared stability
in string theory.

Finally, note that the SUSY we found is a symmetry of the gauge fixed
Lagrangian \v, hence it is respected by interactions, and even non-
perturbative effects (in $1/N$), since it holds for any $N$.

\newsec{The asymptotic spectrum of bound states}

We turn now to a discussion of the eigenvalue problem for the light cone
Hamiltonian \ab. The analysis of section 2 suggests an exponentially growing
density of states $\rho(E)\sim E^a e^{bE}$, and our main task in this
section is to verify this. Again, we will only outline the
derivation of the main results,
postponing details to \dk.

First, it is convenient to rewrite $P^-$ \ab\ in terms of
fermionic oscillators only, using \ac. One finds \kl:
\eqn\pminus{\eqalign{&P^-=(m^2-2g^2N)\int_0^\infty{dk\over k}\psi_{ab}(-k)
\psi_{ba}(k)+\cr
&g^2\int_0^\infty dk_1\cdots dk_4\{A(k_i)\delta(k_1+k_2-k_3-k_4)
\psi_{ab}(-k_3)\psi_{bc}(-k_4)\psi_{cd}(k_1)\psi_{da}(k_2)+\cr
&B(k_i)\delta(k_1+k_2+k_3-k_4)[\psi_{ab}(-k_4)\psi_{bc}(k_3)
\psi_{cd}(k_2)\psi_{da}(k_1)+\cr
&\psi_{ab}(-k_1)\psi_{bc}(-k_2)\psi_{cd}(-k_3)
\psi_{da}(k_4)]\}}}
where
\eqn\AB{A(k_i)={1\over(k_1+k_2)^2}-{1\over(k_1-k_4)^2};\;\;
B(k_i)={1\over(k_1+k_2)^2}-{1\over(k_2+k_3)^2}}
A general wave function of a (single string) bosonic bound state is:
\eqn\psin{\eqalign{\Psi_B&=\sum_{n=1}^\infty\Psi_{2n}\cr
\Psi_n&=\int_0^1\cdots\int_0^1dx_1\cdots dx_n\delta(1-\sum_{i=1}^n
 x_i)\phi_n(x_1,
\cdots,x_n){\rm Tr}[\psi(-x_1)\cdots\psi(-x_n)]|0\rangle\cr}}
The total $p^+$ of \psin\ has been set to 1 by Lorentz invariance.
The extension to fermionic bound states
is obvious. On general grounds \tho, one chooses the ``boundary
conditions" for the functions $\phi_n(x_1,\cdots, x_n)$
(which weigh different distributions of the total momentum $p^+$
between the $n$ constituent quarks):
\eqn\bc{\phi_n(0,x_2,\cdots,x_n)=0}
Note that since the fermions $\psi$ are real, there is a permutation
symmetry of $\phi_n$:
\eqn\perm{\phi_n(x_1,\cdots,x_n)=(-1)^{n-1}\phi_n(x_2,\cdots,x_n,x_1)}
At large $N$, $P^-$ \pminus\ operates within the space $\Psi_B$
(there is no string splitting). Nevertheless, the problem of diagonalizing
$P^-$,
\eqn\diag{P^-\Psi_B=\lambda\Psi_B}
is much more formidable than in the 't Hooft model. The main complication
is due to the terms in $P^-$ which change fermion number. These
couple different $n$'s in \psin.
As a first step, we would like in this section
to find the {\it asymptotic} spectrum of states at high excitation levels.
This is interesting for comparison with section 2 and for developing
a string description of the theory.

Recall the 't Hooft model \tho:
the spectrum of masses (squared) is given by the eigenvalues $\lambda$ of the
integral equation,
\eqn\thooft{m_q^2({1\over x}+{1\over1-x})\phi(x)-\int_0^1{dy\over(x-y)^2}
\phi(y)=\lambda\phi(x)}
where $m_q$ is the renormalized mass of the quark.
A general analytic solution of \thooft\ is unknown, but at large $\lambda$
things simplify; the integral is dominated
by the region $x\simeq y$, corresponding to the long range Coulomb force,
and the contribution of the mass term
is negligible compared to the binding energy. Thus one
can replace \thooft\ by:
\eqn\thoofta{-\int_{-\infty}^\infty{dy\over(x-y)^2}\phi(y)=\lambda\phi(x)}
The eigenfunctions of \thoofta\ are $\phi(x)=\sin px$, with $\lambda
=\pi p$. The boundary conditions $\phi(x=0)=\phi(x=1)=0$ pick out
$p=n\pi$. The asymptotic spectrum is $\lambda_n\simeq\pi^2n$.

It is not difficult to repeat the above considerations for
the case of adjoint matter. For highly excited states $\Psi_B$ in \diag\
(it is clearly enough to consider bosonic bound states because of
the asymptotic SUSY of section 3), the action of $P^-$
on $\Psi_B$ is dominated by singular terms in $P^-$,
corresponding to long range Coulomb forces. Thus the mass term
in \pminus\ can be neglected;
most of the mass of the bound states is due to the binding energy.
Furthermore, since only $A(k_i)$
in \pminus\ contains a singular term, while $B(k_i)$ is regular
(see \AB), for highly excited states we should be able to drop the terms which
produce and annihilate pairs; we will return to their  role
below. The operator one needs to diagonalize
then is:
\eqn\ass{P^-_{c}=
-g^2\int dk_1\cdots dk_4{1\over(k_1-k_4)^2}\delta(k_1+k_2-k_3-k_4)
\psi_{ab}(-k_3)\psi_{bc}(-k_4)\psi_{cd}(k_1)\psi_{da}(k_2)}
which is of interest in its own right, since it summarizes the effect of
the $1+1$ dimensional (confining) Coulomb interaction.
$P^-_{c}$
acts diagonally on the $\Psi_n$ \psin. An easy calculation shows that
\diag\ with \ass\ substituted for $P^-$ gives rise to the
integral equation:
\eqn\eig{{\lambda\over g^2N}\phi_n(x_1,\cdots,x_n)=-
\int_{-\infty}^\infty{dy_1\over(x_1-y_1)^2}
\phi_n(y_1, x_1+x_2-y_1,\cdots,x_n)\pm{\rm cyclic}\;\;{\rm permutations}}
The right hand side is a sum of $n$ cyclic permutations of $(x_1,\cdots,x_n)$
with a $(-1)^{n-1}$ for each elementary cyclic permutation
(because of \perm).

Eq. \eig, which is the analog of \thoofta\ for our case, is significantly
simpler to solve than the original \diag. One can explicitly find the
eigenvalues and eigenmodes $\phi_n$. The general analysis
is still quite involved, therefore here we will write down explicitly
the wave functions $\phi_n$ for the first three non-trivial
cases, and state the general
result for the eigenvalues, leaving the details to \dk. The first non-trivial
case is $n=2$ (approximately, a bound state of two adjoint
quarks). The wave function
$\phi_2$ satisfies \eig:
\eqn\two{{\lambda\over2g^2N}\phi_2(x_1)=-\int{dy_1\over(x_1-y_1)^2}
\phi_2(y_1)}
Since \perm\ $\phi_2(x)=-\phi_2(1-x)$, the solution which respects the boundary
conditions \bc\ is:
\eqn\wavetwo{
\phi_2(x)=\sin\pi n_1x;\;\;\;\lambda_{n_1}=2g^2N\pi^2n_1;\;\;\;
n_1\in 2{\bf Z};\;\;\;n_1>>1}
Consider next $n=4$ (bound states of four adjoint quarks). There are
many solutions of \perm, \eig, but the only one satisfying the boundary
conditions
is:
\eqn\wavefour{\eqalign{
&\phi_4(x_1,\cdots,x_4)=\sin\pi n_1(x_1+x_2)
\sin\pi n_2(x_2+x_3)\cr
&-\sin\pi n_1x_1
\sin\pi n_2 x_3+\sin\pi n_1x_2\sin\pi n_2x_4+(n_1\leftrightarrow n_2)\cr}}
where $n_1, n_2$ are even integers.
This has eigenvalue \diag:
\eqn\eigfour{\lambda_{n_1, n_2}
=2g^2N\pi^2(n_1+n_2); \;\;n_1, n_2\in 2{\bf Z};\;\;\;n_1+n_2>>2}
It is instructive to verify \bc, \perm, \eig\ on \wavefour\
(one must remember to impose $\sum x_i=1$). Note that the wave function
is symmetric under interchange of $n_1, n_2$; there is one state
for each pair $n_1>n_2$.
$\phi_4$ \wavefour\ vanishes when $n_1=n_2$.

As our final example we state without proof the result for six quark states
($n=6$).
\eqn\wavesix{\eqalign{&\phi_6(x_1,\cdots,x_6)=
\sin\pi n_1(x_1+x_2+x_3)\sin\pi n_2(x_2+x_3+x_4)\sin\pi n_3(x_3+x_4+x_5)\cr
&+\sin\pi n_1(x_6+x_1)\sin\pi n_2(x_1+x_2+x_3)\sin\pi n_3(x_3+x_4)\cr
&-\sin\pi n_1(x_1+x_2)\sin\pi n_2(x_2+x_3+x_4)\sin\pi n_3(x_4+x_5)\cr
&+\sin\pi n_1(x_2+x_3)\sin\pi n_2(x_3+x_4+x_5)\sin\pi n_3(x_5+x_6)\cr
&+\sin\pi n_1(x_1+x_2)\sin\pi n_2(x_2+x_3)\sin\pi n_3x_5\cr
&-\sin\pi n_1(x_2+x_3)\sin\pi n_2(x_3+x_4)\sin\pi n_3x_6\cr
&+\sin\pi n_1(x_3+x_4)\sin\pi n_2(x_4+x_5)\sin\pi n_3x_1\cr
&-\sin\pi n_1(x_4+x_5)\sin\pi n_2(x_5+x_6)\sin\pi n_3x_2\cr
&+\sin\pi n_1(x_5+x_6)\sin\pi n_2(x_6+x_1)\sin\pi n_3x_3\cr
&-\sin\pi n_1(x_6+x_1)\sin\pi n_2(x_1+x_2)\sin\pi n_3x_4\cr
&-\sin\pi n_1x_1\sin\pi n_2x_3\sin\pi n_3x_5
+\sin\pi n_1x_2\sin\pi n_2x_4\sin\pi n_3x_6\cr
&-\sin\pi n_1(x_1+x_2+x_3)\sin\pi n_2x_2\sin\pi n_3x_5
+\sin\pi n_1(x_2+x_3+x_4)\sin\pi n_2x_3\sin\pi n_3x_6\cr
&-\sin\pi n_1(x_3+x_4+x_5)\sin\pi n_2x_4\sin\pi n_3x_1+{\rm permutations}
\;\;{\rm of}\;\; (n_1, n_2, n_3)
\cr}}
$n_1$, $n_2$, $n_3$ must again be even integers to ensure
\bc, \perm, \eig. There is one state
(whose wave function is \wavesix) for each choice of $n_1>n_2>n_3$
($\phi_6$ vanishes when any two $n_i$ coincide).
The eigenvalue \diag\ is,
\eqn\eigsix{\lambda_{n_1,n_2,n_3}
=2g^2N\pi^2(n_1+n_2+n_3);\;\;n_i\in 2{\bf Z};\;\;n_1+n_2+n_3>>3}
The pattern  of \wavetwo, \eigfour, \eigsix\ continues for higher
$n$ as well\foot{I am grateful to S. Shenker for important discussions on this
matter.}.
For given $n=2k$ there is one state
satisfying the boundary conditions \bc\ for each choice of
$n_1>n_2>\cdots>n_k$, with the spectrum:
\eqn\eigk{M^2_{n_1,\cdots,n_k}
=4g^2N\pi^2(n_1+n_2+\cdots+ n_k);\;\;\;n_i\in 2{\bf Z};\;\;\;
\sum_1^k n_i>>k}
The corresponding wave functions $\phi_{2k}$ vanish whenever $n_i=n_j$ for any
$1\leq i<j\leq k$. This can be conveniently summarized in a string
motivated parametrization:
$n_1=l_1+\cdots+ l_k$,
$n_2=l_2+\cdots+ l_k$, $\cdots$, $n_k=l_k$. We find non-trivial states
only when all ``excitation numbers'' $l_i>0$, ($i=1,\cdots, k$).

Of course, \eigk\
provides direct confirmation
of an exponentially growing density of states\foot{
The generating functional is ${\rm Tr}\; e^{-sM^2}\simeq
\prod_{n=1}^\infty(1+e^{-sn})$.}, and the existence of
a Hagedorn transition, which was obtained indirectly in section 2
from a different point of view.

So far we have concentrated on the Coulomb part of the light cone Hamiltonian
\ass\ finding a rather rich spectrum of states. Two natural questions arise at
this stage; do the wave functions
$\phi_{2k}$ corresponding to the spectrum \eigk\ form a complete set, and how
do the
particle creation and annihilation
terms
in \pminus\
modify the spectrum?
To discuss these questions, it is useful to introduce a $Z_2$ symmetry of the
theory, $T$:
\eqn\ttt{T:\;\psi_{ab}\to\psi_{ba}}
The Hamiltonian $P^-$, \pminus\ is invariant under $T$; hence its eigenstates
have definite parity under the action of $T$. The wavefunctions $\phi_{2k}$
constructed above can be shown to satisfy $T\phi_{2k}=(-1)^{k+1}\phi_{2k}$.
This is related to their antisymmetry under reflection:
\eqn\sym{\phi_{2k}(x_1,x_2,\cdots, x_{2k})=-
\phi_{2k}(x_{2k},x_{2k-1},\cdots, x_1)}
Thus it appears that we are missing all the eigenstates of $P^-_c$ with
the opposite $Z_2$ charge, and therefore the set $\{\phi_{2k}\}$ is incomplete.
We have been unable to find any eigenfunctions $\tilde\phi_{2k}$ (of $P^-_c$)
with $T=(-1)^k$
which satisfy all the constraints (in particular the boundary conditions),
but it is certainly possible that
there are additional bound states in the theory to the ones exhibited above;
e.g. one may try to relax the boundary conditions;
we will return to this issue in \dk.
Note that the role of particle creation and annihilation can only be studied
once
the sector of the theory with $T=(-1)^k$ is understood; the mixing between
our $\phi_{2k}$ due to pair creation vanishes\foot{
This is in sharp contrast to
what happens for bosonic adjoint matter.},
because of the $Z_2$ symmetry:
$\langle\phi_{2k}|P^-|\phi_{2k\pm2}\rangle=0$.
Physically, the hope is that one can absorb the effects of pair production
in a renormalization of the quarks $\psi_{ab}$. For highly excited states
the only interaction between the renormalized quarks should be the Coulomb
interaction
\ass.

\newsec{Supersymmetry breaking and the extreme infrared theory}

In the previous section we have concentrated on the asymptotic
behavior of the spectrum in the bosonic sector. Assuming SUSY is not
spontaneously broken (for $m^2=g^2N$), the spectrum of fermionic bound states
is the same.
It is time to present a proof of that assumption. We will actually
give two arguments for this result,
the first of which is the following. Consider a bound state $\Psi_B$
\psin. The fermionic state related to it by SUSY is $G_0\Psi_B$. We have
to verify that this state is a physical bound state. Clearly, $G_0\Psi_B$
is an eigenstate of $P^-$; hence, it is enough to check that it satisfies
the boundary conditions \bc. A straightforward calculation \dk\ reveals that
this is indeed the case. As a simple example, consider:
$$G_0\int_0^1dx\phi_2(x){\rm Tr}[\psi(-x)\psi(x-1)]|0\rangle=
2\int_0^1dx\int_0^xdy\phi_2(x){\rm Tr}[\psi(-y)\psi(y-x)\psi(x-1)]|0\rangle$$
This has the form \psin\ with $\phi_3(x_i)=-2\phi_2(x_1)
-2\phi_2(x_2)-2\phi_2(x_3)$, which indeed vanishes for $x_1=0$,
because of the antisymmetry and boundary conditions satisfied by $\phi_2$.
This generalizes for all $n$.
Hence $G_0\Psi_B$ is physical and SUSY is unbroken.

A second way to show that for $m^2=g^2N$ SUSY is not spontaneously
broken is to establish the absence of massless fermions in the spectrum.
Of course, a massless Goldstone fermion must appear if SUSY is to be
spontaneously broken. Since it is clear from \pminus\ that the masses of bound
states are monotonically increasing functions of the bare quark mass $m$,
it is enough to establish this fact for $m=0$.

Consider then \a\ at $m=0$. At very large distance scales the gauge coupling
$g\to\infty$
and the $F_{\mu\nu}^2$ term decouples. The resulting theory is closely related
to the Lagrangian construction of coset models (see e.g. \ref\gaw{
K. Bardacki, E. Rabinovici and B. Saring, \np {\bf B299} (1988) 151;
K. Gawedzki
and A. Kupianen, \np {\bf B320} (1989) 625.})
and is quite well understood. It is
a conformal field theory with central charge $C_{IR}=C_{UV}-C_{gauge}$
where $C_{UV}=\half(N^2-1)$ is the central charge of the free fermion theory,
and $C_{gauge}={N(N^2-1)\over N+N}=\half(N^2-1)$ is the amount by which
gauging the level $N$ $SU(N)$
affine Lie algebra reduces $C$. Thus we find that in the infrared,
$C_{IR}=0$: there are no massless degrees of freedom at large distances.
In particular there are no massless fermions (which would require
$C_{IR}\ge\half$) for $m=0$ and therefore for all $m$. Hence
SUSY can not be spontaneously broken.

The infrared theory
(for massless adjoint quarks):
\eqn\lir{\cl_{IR}=\bar\psi\gamma^\mu D_\mu\psi}
is actually an interesting topological theory in space-time. Although,
as we saw, it contains no field theoretic degrees of freedom,
the theory contains a rich spectrum of global modes, whose correlation
functions are independent of position (since we have taken the distance
scale to infinity). There is a well developed
technology  for studying these states
\gaw: for each primary operator of the fermion $SU(N)$ Kac-Moody algebra
there is one
global mode in the gauged theory \lir\ at ghost number zero\foot{
There are states at other ghost numbers as well; we will not
be concerned with those here.},
obtained
by dressing the fermionic operator with gauge fields; $A_\mu$
does have global degrees of freedom in $2d$.
The group theory
leads to the following result \dk:
for each Young diagram $R$ with $n$ boxes, there is one state
in the infrared theory, obtained by multiplying $n$ $\psi$'s
(the representations of $SU(N)$ that appear in the adjoint fermion
conformal field theory are all representations
of the form $R\times \bar R$). Since there are $P(n)$
such diagrams, we see that in the limit $N\to\infty$ one finds a very
rich spectrum of global modes.

This is to be compared with a similar calculation in the 't Hooft model,
where one finds one state for each completely antisymmetric Young
diagram with $n$ boxes. Thus there is one state for each integer $n$.
It is amusing that in both cases (of adjoint and fundamental matter)
there seems to be a (very) rough correspondence between the spectrum of massive
bound states at finite $g$, and that of the global modes in the extreme IR
limit (in a string description, the nature of the corresponding states
would appear to be quite different).

There has been some recent work on realizing space-time topological theories
as target space theories corresponding to topological world sheet theories
\ref\rab{S. Elitzur, A. Forge and E. Rabinovici, \np {\bf B388} (1992) 131.}.
Perhaps an understanding of the world sheet description of the extreme
IR limit of adjoint QCD
would be a useful first step towards a complete string
description of the theory \a\ at all scales.

\newsec{Concluding Remarks}

The main motivation for this work has been to understand the relation (if any)
of confining
large $N$ gauge theories with a non-trivial spectrum of bound states
and string theory. $2d$ QCD coupled to adjoint matter is perhaps the simplest
theory with an (exponentially) infinite number of Regge trajectories;
it is closely related
to four dimensional QCD via dimensional reduction,  and should provide a
useful laboratory for studying the relation of QCD to string theory.
Many of our results for this theory will look familiar to string theorists,
but (barring an inspired guess) much remains to be done before the
correspondence can be made more precise.

The asymptotic spectrum of infinitely many linear Regge trajectories looks of
course quite stringy. But it is (probably) not exactly linear down to
the ground state. Furthermore, it is interesting that
the spectrum resembles that of an open string (of course, if at all,
this gauge theory is expected to be related to a closed string):
there are $n_i$ \eigk\ but no $\bar n_i$.
In closed superstring theory (a natural candidate would be the three
dimensional version of
\ref\dn{D. Kutasov and N. Seiberg, Phys. Lett. {\bf 251B} (1990) 67.}),
one would generically find states parametrized by
$n_i$, $\bar n_i$, from left and right movers
on the world sheet.
It is also interesting that states with ``occupation numbers" $l_i=0$
don't seem to appear in QCD \eigk.
This reduction of the number of states
of the ``QCD string" compared to a fundamental one may actually be
responsible (among other things) for the absence of space-time gravity in
the former.

The asymptotic supersymmetry discussed in section 3 may also be a manifestation
of a stringy structure, but this issue is far from clear to the author.
We haven't discussed in detail the bosonic version of adjoint QCD, but
it too has a Hagedorn transition (section 2) and an infinite number
of asymptotically linear Regge trajectories \dk. In string theory,
such a system would be unstable in the infrared. The question of whether
bosonic adjoint matter coupled to $2d$ QCD also develops an infrared
instability in the $1/N$ expansion remains open. We hope to return to it
in \dk.

Other favorable signs include the pattern of masses of winding modes
discussed in section 2, and the topological theory in the infrared,
which probably has a world sheet interpretation.
Possible future steps in the attempt to uncover a string description
are a study of the S-matrix of the model, which may provide
more hints (like duality, the exact mass spectrum etc), and a direct
construction of a string in space-time out of the quarks, which act as small
bits of string.
And, if the string
idea fails, adjoint $2d$ QCD seems to be a good place to understand why, and
find alternative ways to deal with the physics of a weakly interacting
system containing an infinite number of linearly rising Regge trajectories.
Clearly, a much better understanding of that physics
is required to address these issues.

\bigbreak\bigskip\centerline{{\bf Acknowledgements}}\nobreak

I am grateful to T. Banks,
P. Ho$\rm {\check r}$ava, I. Klebanov, E. Martinec, N. Seiberg and S. Shenker
for discussions.
This work was partially supported by DOE grant DEFG02-90ER-40560 and
the L. Block foundation.

\listrefs
\bye